\documentclass{article}
\usepackage{graphicx,graphics}
\usepackage{amsmath}
\usepackage{ifpdf}
\usepackage{amsfonts}
\usepackage{amssymb}
\usepackage[spanish]{babel}
\usepackage[latin1]{inputenc}
\newtheorem{theorem}{Teorema}

\title{Acerca del Algoritmo de Dijkstra}

\author{Alvaro H. Salas S.\thanks{Departamento de Matem\'aticas, Universidad de
Caldas, Departamento de Matem\'aticas, Universidad Nacional de
Colombia, Manizales. \emph{email} : asalash2002@yahoo.com}}

\date{}
\begin {document}
\maketitle
\begin{abstract}
In this paper we prove the correctness of Dijkstra's algorithm. We
also discuss it and at the end we show an application. \\En este
artículo realizamos una descripción detallada del algoritmo de
Dijkstra, justificando su correctitud, discutiéndolo y mostrando
algunas de sus aplicaciones.
\end{abstract}
\emph{Palabras claves}: grafo, digrafo, digrafo ponderado, digrafo
pesado, peso de un camino, camino de coste mínimo, camino minimal.
\section{Introduccción}
Dado un grafo con etiquetas no negativas, se trata de calcular el
coste del camino mínimo desde un vértice dado al resto (ing.,
single-source shortest paths). La utilidad de un procedimiento que
solucione esta cuestión es clara: el caso más habitual es disponer
de un grafo que represente una distribución geográfica, donde las
aristas den el coste (en precio, en distancia o similares) de la
conexión entre dos lugares y sea necesario averiguar el camino más
corto para llegar a un punto partiendo de otro (es decir, determinar
la secuencia de aristas para llegar a un nodo a partir del otro con
un coste mínimo). La solución más eficiente a este problema es el
denominado algoritmo de Dijkstra, en honor a su creador, E.W.
Dijkstra. Formulado en 1959 en "A note on two problems in connexion
with graphs", Numerical Mathematica, 1, pp. 269-271, sobre grafos
dirigidos, el algoritmo de Dijkstra  es un algoritmo voraz
(algoritmo goloso) que genera uno a uno los caminos de un nodo $a$
al resto por orden creciente de longitud; usa un conjunto $S$ de
vértices donde, a cada paso del algoritmo, se guardan los nodos para
los que ya se sabe el camino mínimo y devuelve un vector indexado
por vértices, de modo que para cada uno de estos vértices podemos
determinar el coste de un camino más económico (de peso mínimo)  de
$a$ a tales vértices. Cada vez que se incorpora un nodo a la
solución, se comprueba si los caminos todavía no definitivos se
pueden acortar pasando por él.
\section{Preliminares} Recordemos que un digrafo (eng. directed
graph) es una pareja $G=(V,E)$, en donde $V$ es un conjunto  y $E$
es una relación binaria irreflexiva sobre $V$, es decir, un
subconjunto de $V\times V$ tal que $(x,x)\notin E$ para todo $x\in
E.$ Decimos que $V$ es el conjunto de vértices y que $E$ es el
conjunto de aristas (eng. edges). En los sucesivo supondremos que
$V$ es finito. Dada una arista $(u,v)$, decimos que los vértices $u$
y $v$ son adyacentes y que la arista es incidente en ellos $(u,v)$.
Se llama camino de $u$ a $v$, denotado por $u-v$, a toda sucesión
finita de vértices
\begin{equation}\label{eq01}
 u=x_0,x_1,\ldots,x_i,x_{i+1},\ldots,x_{m-1},x_m=v,
\end{equation}
\ifpdf
\begin{figure}[h]
\begin{center}
\includegraphics[width=13cm]{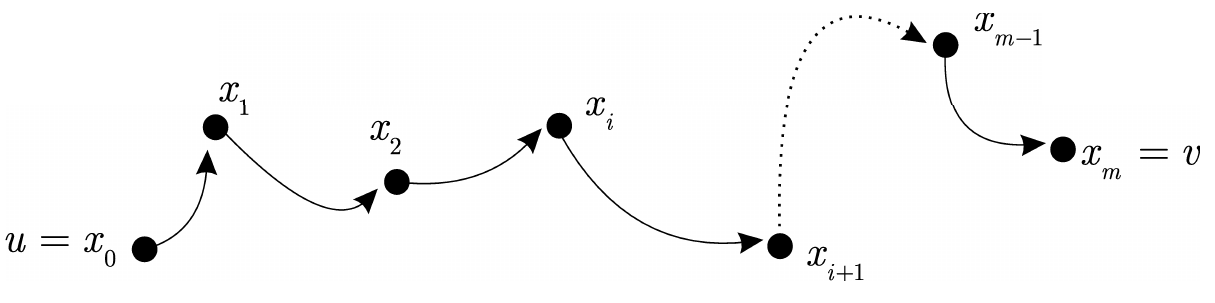}\\
\caption{Un camino $u-v$}\label{fig1}
\end{center}
\end{figure}
\else \fi   \noindent de modo que $(x_i,x_{i+1})$ es una arista para
cada $i=0,1,\ldots,m-1$. Por definición, (\ref{eq01}) es un camino
de longitud $m$, $u$ es su vértice inicial, $v$ es su vértice
terminal y
$x_1$, $x_2$,..., $x_{m-1}$ son sus vértices internos.\\
Un digrafo es ponderado o pesado, si a cada arista $(u,v)$ se le
asigna un número real, denotado por $p(u,v)$ y  llamado su peso. En
lo sucesivo consideraremos un digrafo ponderado $G$ con pesos
positivos : $p(u,v)>0$ para toda arista $(u,v)\in E.$ El peso del
camino (\ref{eq01}) se define como la suma de los pesos de sus
aristas :
\begin{equation}\label{eq02}
p(u-v)=p(x_0,x_1)+p(x_1,x_2)+\cdots+p(x_{m-1},x_m).
\end{equation}
Un camino es de coste mínimo si no es posible encontrar ningún
camino de $u$ a $v$ cuyo peso sea menor que $p(u-v)$. En otras
palabras, dado cualquier camino de $u$ a $v$
$$ u=y_0,y_1,\ldots,y_{k-1},y_k=v,$$
se debe tener
$$p(u,y_1)+p(y_1,y_2)+\cdots+p(y_{k-1},v)\geq p(u-v).$$
\section{El algoritmo de Dijkstra. Descripción y sustentación de su correctitud.}
Sea $G=(V,E)$ un digrafo ponderado con pesos positivos de $n$
vértices. Supongamos que $a$ y $z$ son dos vértices en $V$, de modo
que $z\neq a$ y existe al menos un camino de $a$ a $z$. Nuestro
principal objetivo consiste en hallar un camino $a-z$ de coste
mínimo. Este problema se resuelve de manera eficiente mediante el
algoritmo de Dijkstra. El algoritmo inicia en el vértice $a$ y
construye un camino de coste mínimo
$$a=u_0,u_1,\ldots,u_{m-1},u_m=z,$$
tal que $a-u_i$ es un camino de coste mínimo para cada
$i=0,1,\ldots,m$.
\subsection{Algoritmo de Dijkstra}
\noindent \emph{\textbf{Entrada}}: Grafo ponderado dirigido de $n$
vértices con pesos positivos; $a$ y $z$ vértices distintos tales que
existe algún camino de
$a$ a $z$.\\
\emph{\textbf{Salida}}: Peso de un camino de coste mínimo de $a$ a
$z$.
\begin{itemize}
\item \textbf{\emph{Paso 1}} : Definimos $S_0=\emptyset$,
$T_0=V$. Asignamos a cada vértice $v$ en $V$ una etiqueta (eng.
label) como sigue : $L(v)=0$ si $v=a$ y $L(v)=\infty$ para $v\neq
a$.
  \item \textbf{\emph{Paso 2}} :  Para $i=1,2,\ldots,n$ :
Supongamos que hemos construido los conjuntos $S_0$, $S_1$,...,
$S_{i-1}$. Hacemos  $T_{i-1}=V\setminus S_{i-1}$. Si $z\in S_{i-1}$,
definimos $S=S_{i-1}$ y detenemos la construcción. En caso
contrario, escogemos el primer vértice $u$ en $T_{i-1}$ con la menor
eqtiqueta,
  es decir,
  $$L(u)=\min\{L(v)\,|\,v\in T_{i-1}\}.$$

  Definimos $u_{i-1}=u$, $S_i=S_{i-1}\cup
  \{u_{i-1}\}=\{u_0,u_1,\ldots,u_{i-1}\}$ (decimos que $u$ entra),
  $T_i=V\setminus S_{i}$ y para cada vértice $v$ en $T_{i}$
  adyacente a $u$ cambiamos su etiqueta $L(v)$ por la nueva etiqueta
  $\min\{L(v),L(u)+p(u,v)\}$ :
$$L(v)\leftarrow \min\{L(v),L(u)+p(u,v)\},$$
es decir, actualizamos la etiqueta de los "vecinos" de $u$ por fuera
de  $S_i$.
\item\textbf{\emph{Paso 3}} : Si $i=n$, definimos $S=S_{n}$ y
 nos detenemos. Si $i<n$, hacemos $i=i+1$ y vamos al Paso 2.
\end{itemize}
El algoritmo de Dijkstra termina en el momento en que encontramos el
primer índice $m$ para el cual $z\in S_m$. En ese momento, $S=S_m$.
\subsection{Correctitud del algoritmo de Dijktra}
La justificación de la correctitud de este notable  algoritmo se
basa en el siguiente
\begin{theorem} Al finalizar la ejecución del algoritmo de Dijkstra
tenemos que, para todo $u\in S$ con $u\neq a$,
\begin{itemize}
 \item[\textbf{A}.] Cualquier camino $a-u$ tiene peso al menos $L(u)$ : $L(u)\leq p(a-u)$
 \item[\textbf{B}.] Existe un camino de $a$ a $u$ de peso igual a
 $L(u)$. En particular, $L(u)$ es finita : $L(u)<\infty.$
\end{itemize}
\end{theorem}
\textbf{\emph{Demostración}}. Haremos la demostración por inducción
sobre $i\geq 1$. Sea $i=1$. Inicialmente (esto es, antes del Paso 2)
$S_0=\emptyset$, $T_0=V$. Es claro que $z\notin S_0$. De acuerdo al
Paso 2, escogemos el vértice $a$ por cuanto su etiqueta es la menor
(los demás vértices tienen etiqueta $\infty>0$), así que el vértice
$a=u_0$ entra y $S_1=\{u_0\}$. Entonces $T_1=V\setminus S_1$. Por
hipótesis, existe al menos un camino de $a$ a $z$. Sea
\begin{equation}\label{eq05}
a=x_0,x_1,\ldots,x_{m-1},x_m=z
\end{equation}
dicho camino. Entonces $x_1\in T_1$ es un vecino de $u_0$, así que
el conjunto de los vecinos de $u_0$ por fuera de $S_0$ es no vacío.
De acuerdo al Paso 2, actualizamos las etiquetas de estos vecinos.
Sea $v$ cualquiera de ellos. Su nueva etiqueta es
$$\min\{L(v),L(a)+p(a,v)\}=\min\{\infty,0+p(a,v)\}=p(a,v).$$
Según esto, la etiqueta de $v$ (que en este momento es $\infty$) se
cambia por el peso de la arista $(a,v)$. Supongamos que $u_1$ es un
vecino con la menor etiqueta (el vecino más económico o más cercano
a $u_0=a$; puede haber más de uno, en cuyo caso podemos escoger
\emph{cualquiera} de ellos), esto es,
$$L(u_1)=\min\{p(a,v)\,|\,v\in T_{1},\,\text{siendo}\,\,v\,\,\text{un vecino de}\,\,a\}=p(a,u_1).$$
 Sea (\ref{eq05})
cualquier camino $a-z$. Tenemos:
$$L(u_1)\leq p(a,x_1)\leq p(a,x_1)+p(x_1,x_2)+\cdots+p(x_{m-1},z)=p(a-z),$$
de modo que la condición \textbf{A} se cumple para $i=1$. De otro
lado, \textbf{B} es verdadera, ya que $a,u_1$ es un camino $a-u_1$
de peso
$p(a-u_1)=p(a,u_1)=L(u_1)$.\\
De acuerdo con los anteriores argumentos, el teorema es válido para
$i=1$.\\  Supongamos que este teorema se cumple para todos los
vértices $u_i\in S$ con $1\leq i\leq k$. Veamos que este teorema
también es válido para
el vértice $u_{k+1}.$\\
Consideremos \emph{cualquier} camino de $a$ a $v=u_{k+1}$ (ver
Figura \ref{fig2}). Veamos que $L(v)\leq p(a-v)$. Si esto fuera
falso, se tendría $L(v)>p(a-v)$. \ifpdf
\begin{figure}[h]
\begin{center}
\includegraphics[height=4cm]{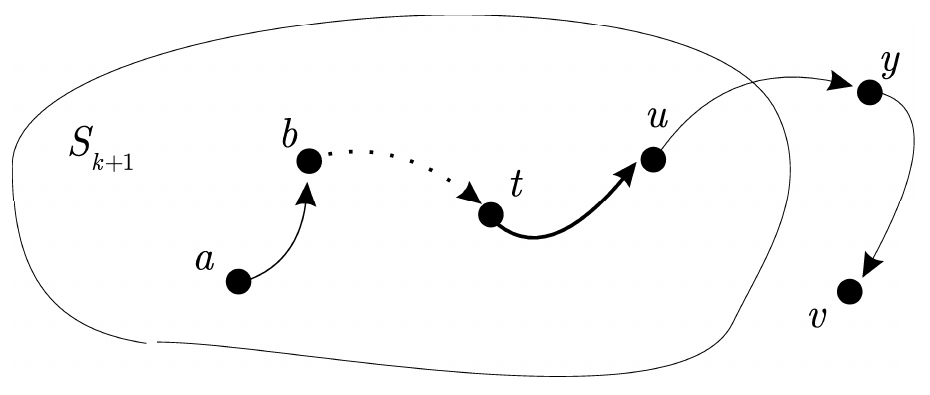}\\
\caption{Un camino arbitrario $u-v$}\label{fig2}
\end{center}
\end{figure}
\else \fi    Supongamos que $y$ es el primer vértice de este camino
por fuera de $S_{k+1}$ y sea $u$ el predecesor de $y$ en dicho
camino. Entonces $u\in S_{k+1}=\{u_0,u_1,\ldots,...,u_k\}$. De
acuerdo con nuestra hipótesis inductiva, $L(u)\leq p(a-u)$. Además,
$L(v)=L(u_{k+1})\leq L(y)$, luego
$$L(v)\leq L(y)\leq L(u)+p(u,y)\leq p(a-u)+p(u,y)=p(a-y)\leq p(a-v)<L(v),$$
lo cual es una contradicción. En consecuencia, la parte \textbf{A}
del teorema se cumple para $i=k+1$. \\
Supongamos que en el camino (\ref{eq05}) el vértice $y=x_j$ es el
primero que se encuentra por fuera de $S_{k+1}=
\{u_0,u_1,\ldots,u_{k}\}$. Ver Figura \ref{fig3}.
 \ifpdf
\begin{figure}[h]
\begin{center}
\includegraphics[height=4cm]{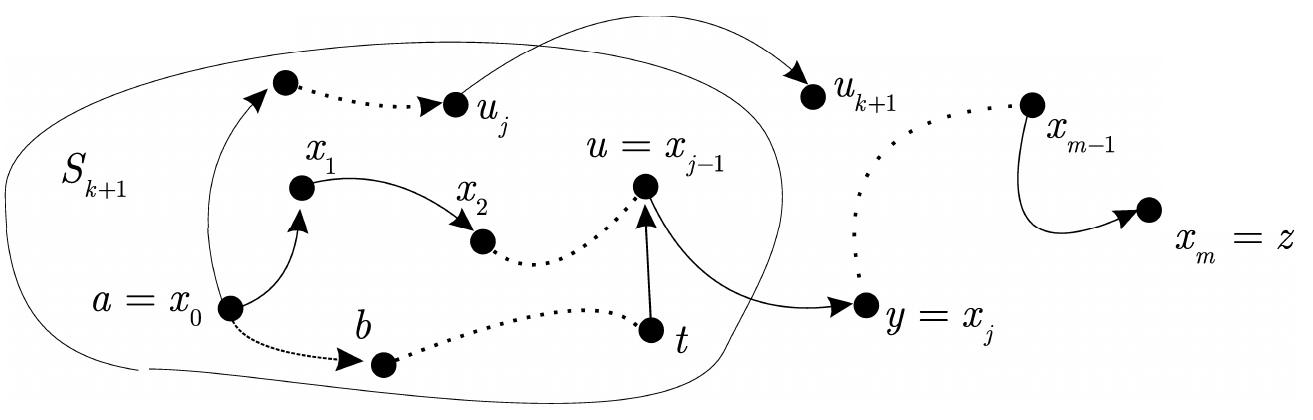}\\
\caption{Existencia de un camino de $a$ hacia $u_{k+1}$ de peso
$L(u_{k+1})$}\label{fig3}
\end{center}
\end{figure}
\else \fi    Entonces $y=x_j$ es vecino de $u=x_{j-1}\in S_{k+1}$.
Por hipótesis inductiva, existe un camino $a,b,\ldots,t,u$ de $a$ a
$u$ de peso $L(u)$, luego $L(u)$ es finito. Cuando $u$ entró a $S$,
la etiqueta de su vecino $y$ fue actualizada, de modo que $L(y)\leq
L(u)+p(u,y)<\infty$. Ahora, ambos, $y$ y $u_{k+1}$, están por fuera
de $S_{k+1}$, así que $L(u_{k+1})\leq L(y)<\infty.$ Inicialmente
(esto es, antes del Paso 2), el vértice $u_{k+1}$ tenía etiqueta
$\infty$ (observemos que $u_{k+1}\neq a$, ya que $a=u_0$ y $k+1\neq
0$). En el momento actual, por lo que acabamos de demostrar,  este
vértice tiene etiqueta finita. Esto significa que en algún momento
fue actualizada y cambió de $\infty$ a $L(u_j)+p(u_j,u_{k+1})$ para
algún $j$ con $j\leq k$. Sin pérdida de generalidad, podemos suponer
que este es el valor actual de $L(u_{k+1})$, es decir,
\begin{equation}\label{eq06}
L(u_{k+1})=L(u_j)+p(u_j,u_{k+1}).
\end{equation}
De acuerdo a nuestra hipótesis inductiva, existe un camino de
$a-u_j$ de peso $L(u_j)$. Agregando a este camino la arista
$(u_j,u_{k+1})$ obtenemos un camino de peso
$L(u_j)+p(u_j,u_{k+1})=L(u_{k+1})$, de modo que \textbf{B} también
se satisface para $i=k+1$. Teorema demostrado. $\blacklozenge$\\\\
Del teorema antwerior se sigue que :
\begin{itemize}
\item[\textbf{a.}] \emph{El algoritmo termina.}\\
En efecto, probemos por inducción sobre $i\geq 1$ que los elementos
de  $S_i$ son distintos.\\
Para  $i=1$, $S_1=\{u_0\}$ y en este caso no hay nada que demostrar.
Supongamos que todos los elementos de $S_i$ son distintos para algún
$i\geq 1$, de modo que $z\notin S_i$. Sea $x$ el último vértice en
$S_{i}\subseteq S$ de un camino $u-z$ y $y$ su vecino. Cuando $x$
entró a $S$, la etiqueat de $y$ se actualizó, luego  $L(y)\leq
L(x)<\infty$. Por lo tanto, es posible escoger en $T_i=V\setminus
S_i$ un elemento con etiqueta mínima finita $\leq L(y)$. Uno de
ellos es precisamente el elemento $u_{i}$. Entonces definimos
$S_{i+1}=\{u_0,u_1,\ldots,u_{i-1},u_i\}$. Es claro que todos los
elementos de $S_{i+1}$ son distintos.\\
Por consiguiente, $i\leq n$ y el algoritmo debe terminar.
\item[\textbf{b.}] \emph{$L(u)$ es el peso de un camino de
coste mínimo de $a$ a $u$ para todo $u\in S\setminus\{a\}$.}\\
 En efecto, puede haber más de un
camino de coste mínimo. Sin embargo, es fácil ver que todos ellos
tienen el mismo peso. Si consideramos uno de estos caminos, su peso
$q$  es menor o igual que el peso del camino de $a$ a $u$ cuya
existencia se garantiza en la parte \textbf{B} del teorema, esto es,
$q\leq L(u)$. De otro lado, según la parte \textbf{A} de este
teorema, todo camino $a-u$ tiene peso al menos $L(u)$, esto es,
$q\geq L(u)$. Concluimos que $q=L(u)$. En particular, tomando
$u=z\in S$, obtenemos que $L(z)$ es el peso de un camino de coste
mínimo de $a$ hacia $z$.
\item[\textbf{c.}] \emph{Si $u_m=z$, entonces}
$$L(u_0)\leq L(u_1)\leq\cdots\leq L(u_j)\leq L(u_{j+1})\leq\cdots \leq L(u_m).$$
En efecto, antes de que los elementos $u_j$ y $u_{j+1}$ entren a
$S=S_m$, éstos se encuentran en $T_{j}=V\setminus S_j$. Se escoge
$u_j$ por cuanto este es uno de los elementos con etiqueta mínima,
luego $L(u_j)\leq L(u_{j+1})$.
\end{itemize}
\section{Discusión}
Supongamos que aplicamos el algoritmo de Dijkstra empezando en un
vértice arbitrario $a$. En calidad de $z$ tomamos \emph{cualquier}
vértice distinto de $a$. Si existe un camino de $a$ a $z$, el
algoritmo funciona. Supongamos que no existe tal camino. En este
caso, $z$ no pertenece a ninguno de los conjuntos $S_i$. Al aplicar
los pasos del algoritmo, nos detenemos en el momento en que todas
las etiquetas de los vértices que aún no han entrado sea $\infty$.
Por lo tanto, dados dos vértices arbitrarios distintos $a$ y $z$, al
tomar $a$ como vértice inicial, si ejecutamos los pasos del
algoritmo, deteniéndonos en el momento en que
$$\min\{L(v)\,\,|\,\,v\in T_i\}=\infty,$$
obtenemos un conjunto $S\subseteq V$ de vértices. Si $z\notin S$,
entonces no hay ningún camino $a-z$. En particular, si $S=\{a\}$,
entonces no existe ninguna arista que parta de $a$, es decir, $a$ no
tiene vecinos. En este caso decimos que $a$ es un vértice aislado.
Si $S=V$, entonces existe un camino de $a$ a cualquier otro vértice
y las etiquetas finales de dichos vértices nos dan un camino de peso
mínimo desde $a$.\\
Cabe anotar que el peso de un camino de coste mínimo de $a$ a $z$
puede ser distinto del peso de un camino de coste mínimo de $z$ a
$a$. Por ejemplo, $a$ y $z$ pueden representar ciudades vecinas.
Digamos que  el peso de las aristas $(a,z)$ y $(z,a)$ representa el
coste de transportar cierta mercancía de una ciudad a otra. En
general, $p(a,z)\neq p(z,a)$. El algoritmo de Dijkstra nos
proporciona $L(z)=p(a,z)$ al tomar $a$ como vértice inicial y
$L(a)=p(z,a)$ si $z$ es el vértice
inicial.\\
En el caso en que en el grafo $G=(V,E)$ sea una relación simétrica,
decimos que el grafo es no dirigido y  las aristas son
bidireccionales ( sobre las aristas no colocamos ninguna flecha ) y
en este caso $L(a)=L(z)$ para cualquier par de vértices del
grafo. \\
\section{Algoritmo de Dijkstra modificado}
En algunas ocasiones nos interesa conocer no solamente el peso de un
camino de coste mínimo entre dos vértices dados, sino también la
ruta a seguir para llegar de uno de estos dos vértices al otro. En
este caso a cada vértice distinto de $a$ le agregamos una etiqueta
adicional, es decir, a cada vértice $v\neq a$ le asociamos una
pareja $\overline{L}(v)$ de la siguiente manera. En el Paso 1,
$\overline{L}(v)=(\infty,\star)$. Si al ejecutar el Paso 2 éste
vértice es vecino de $a$, entonces su nueva etiqueta es
$\overline{L}(v)=(p(u,v),u).$ En caso contrario, esta etiqueta
inicial es $(\infty,\star)$. Supongamos que en determinado momento
la etiqueta de cierto vértice $v\notin S$ se actualiza, es decir que
$v$ es un vecino de un vértice $u$ que acaba de entrar a $S$, de
modo que $L(u)+p(u,v)<L(v)$. Entonces la etiqueta de $v$ se cambia
por $\overline{L}(v)=(L(u)+p(u,v),u).$ El vértice $u$ es predecesor
de un camino de coste mínimo de $a$ a $v$. De esta manera, si al
terminar el algoritmo la etiqueta de $v$ es
$\overline{L}(v)=(L(v),u)$, entonces observamos la etiqueta de $u$.
Sea $\overline{L}(u)=(L(u),t)$. Entonces un camino de coste mínimo
de $a$ a $v$ debe ser de la forma $a,\ldots,t,u,v.$ Al observar la
etiqueta de $t$, obtenemos otro predecesor. Procedemos de esta
manera hasta llegar a un vértice $b$ cuya etiqueta sea
$\overline{L}(b)=(L(b),a)$. En ese momento habremos obtenido un
camino $a,b,\ldots,t,u,v$ de coste mínimo igual a $L(v).$\\

\section{Aplicación}
En esta sección mostraremos una aplicación del algoritmo de Dijkstra
en un caso concreto. La Figura \ref{fig4} muestra un grafo no
dirigido ponderado. Hallemos un camino de coste mínimo de $a$ a $z$.

\ifpdf
\begin{figure}[h]
\begin{center}
\includegraphics[height=4cm]{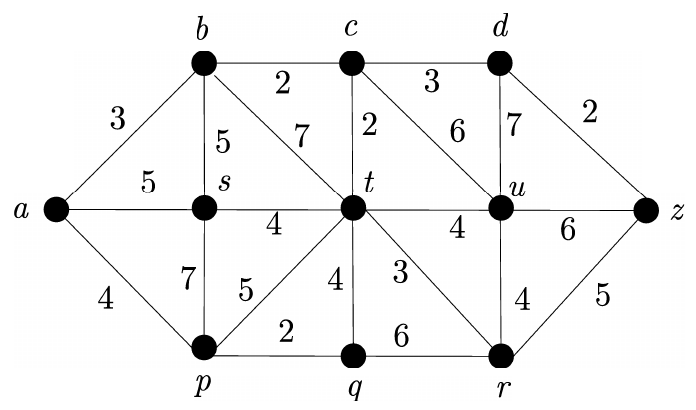}\\
\caption{Grafo ponderado.}\label{fig4}
\end{center}
\end{figure}
\else \fi   La ejecución del algoritmo se ilustra a continuación :

\ifpdf

\begin{center}
\includegraphics[width=17cm]{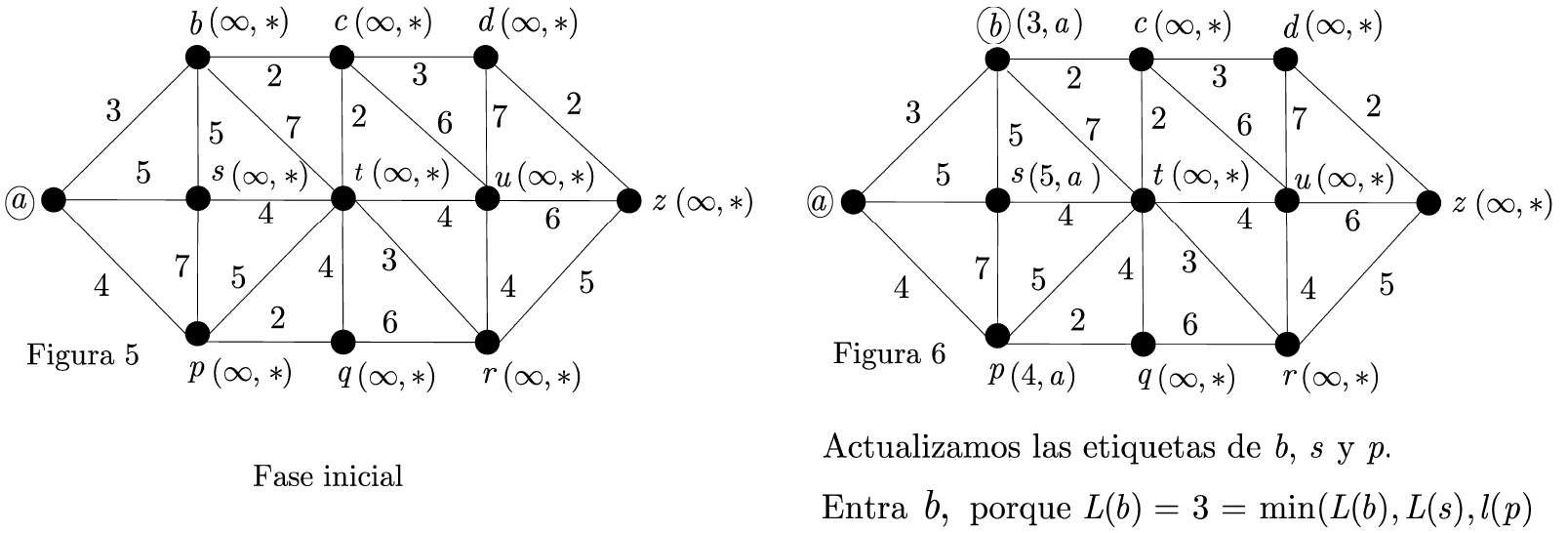}\\
\end{center}

\begin{center}
\includegraphics[width=17cm]{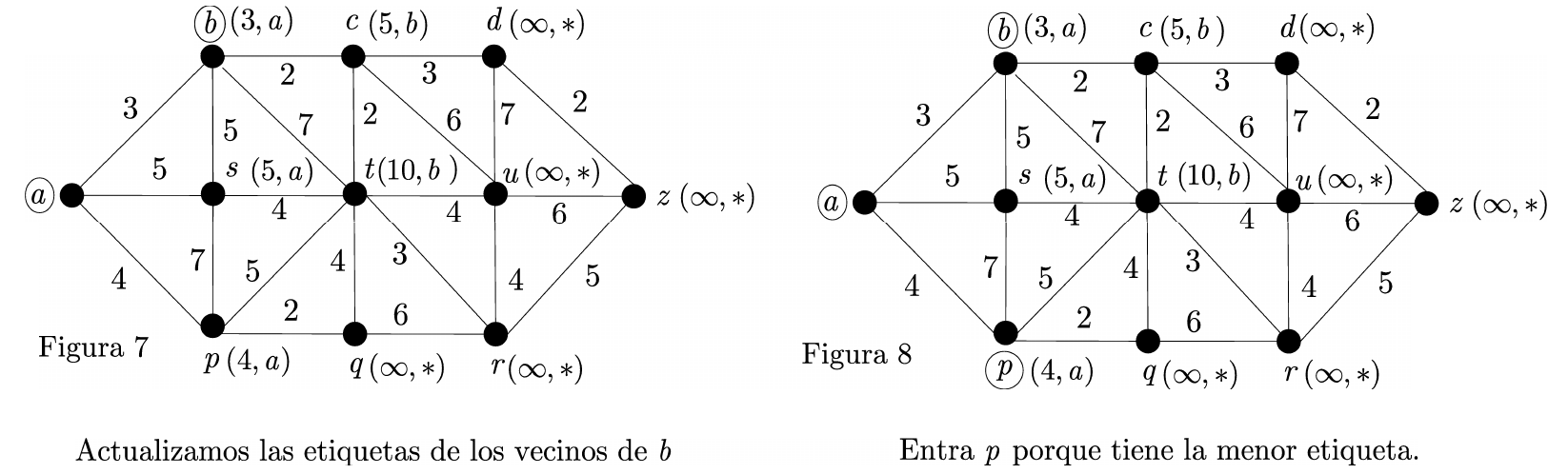}\\
\end{center}

\begin{center}
\includegraphics[width=17cm]{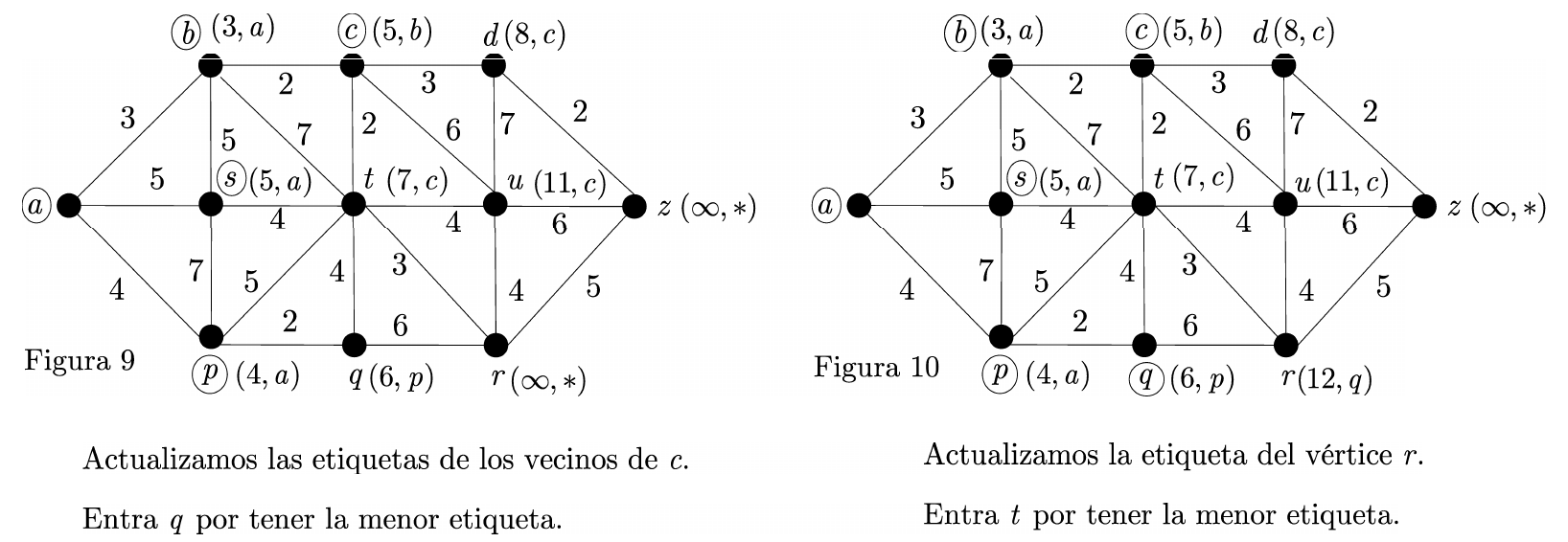}\\
\end{center}

\begin{center}
\includegraphics[width=17cm]{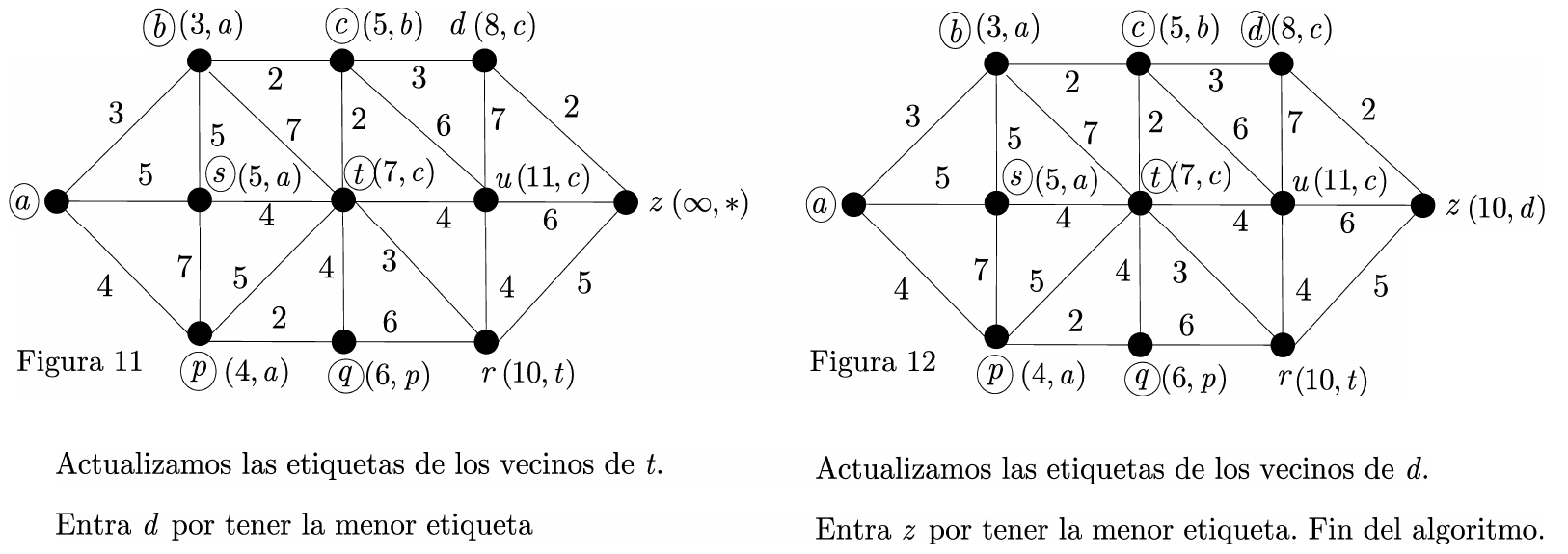}\\
\end{center}
\else \fi De acuerdo a la Figura 12, un camino de coste m{\'i}nimo
de $a$ a $z$ es $a,b,c,d,z$. Este camino tiene peso igual a $10$.


\bigskip
\begin{thebibliography}{99}
\small
\bibitem{baugh}{\sc R. Johnsonbaugh}, \emph{Matemáticas Discretas},
 Prentice Hall, Cuarta Edición, pag. 338--343.
\end{thebibliography}
\end{document}